\documentclass[fleqn,10pt]{wlscirep}
\usepackage[utf8]{inputenc}
\usepackage[T1]{fontenc}
\usepackage{nccmath}

\title{Electro-Quasistatic Animal Body Communication for Chronic Untethered Rodent Biopotential Recording }

\author[1*]{Shreeya Sriram}
\author[1]{Shitij Avlani}
\author[2,3]{Matthew P Ward}
\author[1]{Shreyas Sen}

\affil[1]{School of Electrical and Computer Engineering, Purdue University, West Lafayette, IN, 47906, USA}
\affil[2]{Weldon School of Biomedical Engineering, Purdue University, West Lafayette, IN, 47906, USA}
\affil[3]{Indiana University School of Medicine, Indianapolis, IN, 46202, USA}

\affil[*]{sriram11@purdue.edu, shreyas@purdue.edu}

\begin{abstract}

Continuous multi-channel monitoring of biopotential signals is vital in understanding the body as a whole, facilitating accurate models and predictions in neural research. The current state of the art in wireless technologies for untethered biopotential recordings rely on radiative electromagnetic (EM) fields. In such transmissions, only a small fraction of this energy is received since the EM fields are widely radiated resulting in lossy inefficient systems. Using the body as a communication medium (similar to a 'wire') allows for the containment of the energy within the body, yielding order(s) of magnitude lower loss than radiative EM communication. In this work, we introduce Animal Body Communication (ABC), which utilizes the concept of using the body as a medium into the domain of chronic animal biopotential recording. This work, for the first time, develops the theory and models for animal body communication circuitry and channel loss. Using this theoretical model, a sub-inch$^3$, custom-designed sensor node is built using off the shelf components which is capable of sensing and transmitting biopotential signals, through the body of the rat at significantly lower powers compared to traditional wireless transmissions. \textit{In-vivo } experimental analysis proves that ABC successfully transmits acquired electrocardiogram (EKG) signals through the body with correlation accuracy >99\% when compared to traditional wireless communication modalities, with a 50x reduction in power consumption.

\end{abstract}
\begin{document}

\flushbottom
\maketitle
%
%
\thispagestyle{empty}

\noindent 

\section*{Introduction}

Chronic monitoring of biopotential signals has paved the way for a better understanding of neural pathways along with improved therapeutic treatments.  Recent proliferation in small form-factor wearables has enabled a new domain of continuous health monitoring. This coupled with miniaturized biological sensors, both in the wearable and the implantable domain has resulted in gathering valuable information regarding the body. Surface biopotential signals such as EKG, sEMG (Surface Electromyography), EEG (Electroencephalogram) have been studied as a means of understanding the behavior of the body. Neural recording systems interfaced with the peripheral nervous system have been extensively explored as a method to acquire meaningful data that is used to predict and understand the motor, sensory, proprioceptive, and feedback functions of the brain. 

The use of animals in biological research and medicine has been a longstanding practice given the similarity between the animal and human anatomy and physiology. The current state of the art in animal signal recording includes miniaturized implantable devices with stimulation and recording capabilities. These devices implanted inside the animal body, transmit data to an external unit capable of receiving and processing this information.  A vast majority of animal recording systems still rely on tethered units especially in cases when continuous long-term information is a priority. Tethered systems are not limited by the data rates and are a gold standard for reliable comprehensive information, this, however, does come with a few caveats. Tethered systems are limited by the bias and irritation introduced by these devices on the subject. Long experimental duration is possible, however, the experimental arena is hindered by the need for long wires which in turn results in restricted movement of animals. Noisy systems result from the animals tucking and biting at the wires. Tethered systems require signal conditioning electronics to be placed external to the body, long wires also act as a site for infection and require buffers to prevent signal attenuation\cite{harrison2006low}. To reduce these effects, wireless telemetry of signal information is needed. Wireless recording systems have since evolved from discrete modules to system on chip devices. 
These small form-factor wireless devices eliminate the bias introduced by the tethered systems; however, it is limited by the high-power consumption due to the need for up-conversion of the baseband signal to higher radio frequencies \cite{liu2016fully} and loss due to radiative communication. Animal studies, in particular, require the animal to wear a heavy battery pack to meet this high-power requirement. The need for constant replacement of batteries limits the experimental duration and causes undue stress on the animals which in turn corrupts the data \cite{szuts2011wireless}. On the other hand, perpetual recording could be achieved through wireless power transfer\cite{lee2018implantable}, however, due to the inherent high-power consumption of the sensing node with electromagnetic communication, high-power needs to be harvested, increasing the on-device harvester size significantly.

\begin{figure}
\centering
\includegraphics[width=\linewidth]{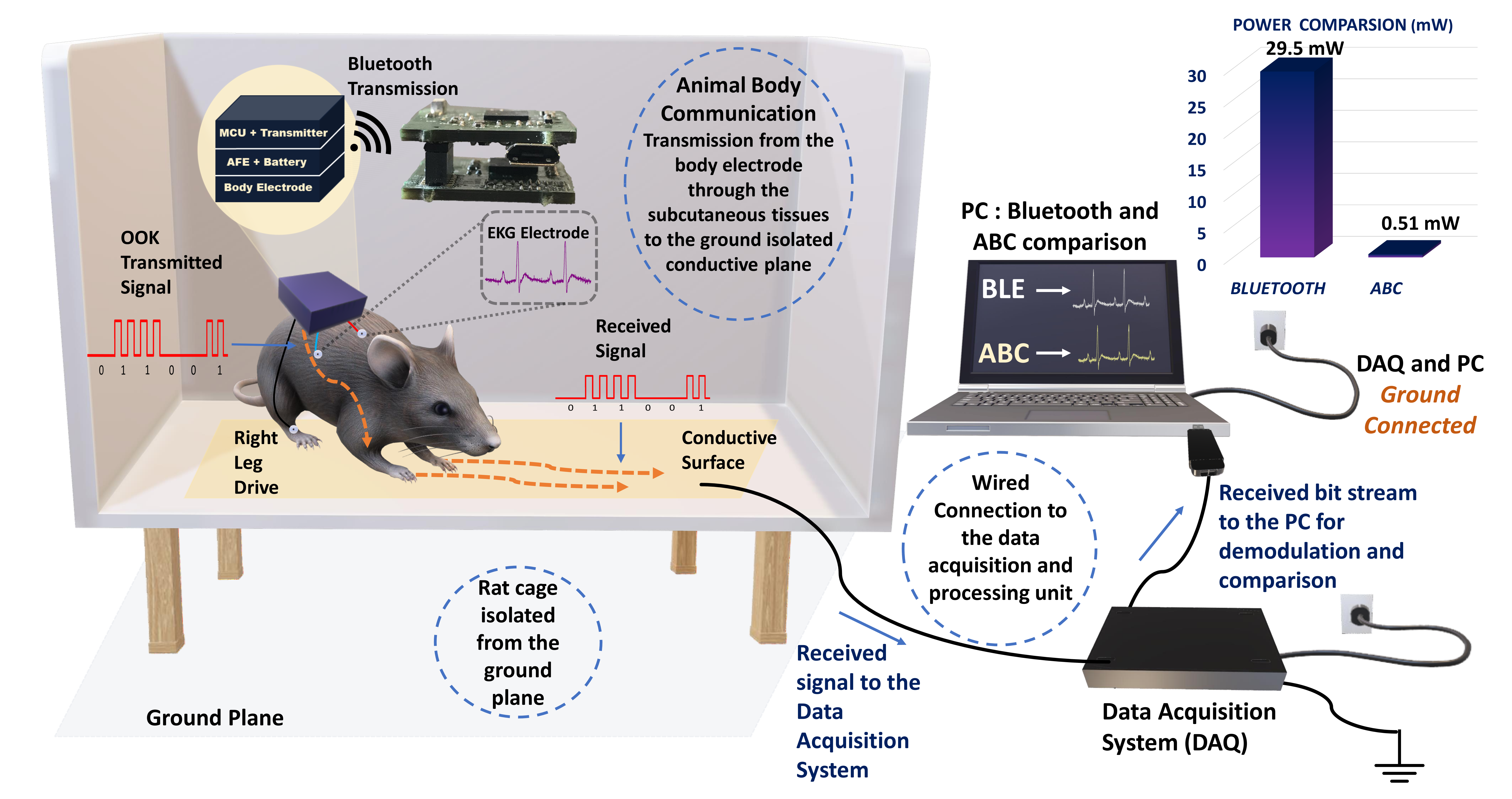}
\caption{\textbf{Animal Body Communication}: a) Overview of Animal Body Communication on a Rodent Model. Custom designed sensor node is placed on the back of the rat. This sensor node is capable of sensing and transmitting the surface biopotential signals via Bluetooth and Animal Body Communication. The sensed signal is transmitted through the body to the conductive surface in the form of OOK (On-Off Keying) sequences. The specially designed rat cage is isolated from the ground surface. A conductive surface is placed on the base of the rat cage which is then connected to a Data Acquisition System (DAQ) which receives the transmitted signals. The Bluetooth receiver and DAQ are connected to a PC for processing, with the DAQ and PC ground referenced.  In this model Bluetooth communication acts as a validity check for ABC. \textit{The rat model in a) was created using Paint 3D.}}
\label{fig:Figure1ABC}
\end{figure}

To overcome these constraints there is a need for a low power communication modality capable of withstanding the continuance of the experiment, along with having an unbiased model in which the animals are free to move in their natural environments. In this work, we introduce a novel communication modality that uses the animal body as a medium to transmit information. This system eliminates the bias introduced by the tethered systems and has a significant size, weight, area, and power benefits compared to electromagnetic communication systems. We demonstrate this \textbf{Electro-Quasistatic Animal Body Communication (EQS-ABC)} as a low loss, efficient channel which addresses the aforementioned drawbacks of both tethered and wireless EM communication systems. Using the body of the animal as a \lq wire\rq    ensures signal confinement resulting in significantly lower losses when compared to traditional wireless technologies. Here, we demonstrate ABC using a rodent model and explain the theory and biophysical models of ABC, followed by ABC demonstration with  a sub-inch$^3$, custom-designed sensor node in the subsequent sections.  Figure \ref{fig:Figure1ABC}, describes the concept of the ABC setup, surface biopotential signals are acquired by a custom-designed sensor node that then transmits the signal using ABC through the subcutaneous tissues of the animal body using EQS-ABC. These signals are picked up by a receiver connected to the ground isolated conductive surface. In this setup, we also transmit the signals using Bluetooth as a method to compare the ABC transmitted signal with an established communication modality. In this pilot study, we use the concept of body communication and extend it into the animal domain enabling long term benefits in energy consumption along with size, weight, and area benefits. The low power requirement enables the use of smaller batteries or coils in the case of energy harvested nodes. We show the concept of ABC applied to animal biomedical studies, this modality can be extended into the neuroscience domain, with implantable nodes to acquire the signals and then communicate using the body as a medium in the future. This work presents the first validation of ABC as a communication modality. Experiments were performed with EKG signals of the rat as the chosen surface biopotential signal. 

\paragraph{}	
\fontsize{11}{12} \textbf{ State of the Art in Wireless Biopotential Recording} 
\paragraph{}

The first biopotential discovery dates back to 1666 when Francesco Redi measured the EMG from a specialized muscle in the electric eel\cite{reaz2006techniques}, the field of animal biopotential recordings evolved from tethered systems to wireless systems in the year 1948 when Fuller and Gordon first used radio communication for biopotential signal transmission \cite{fuller1948radio}. Presently, multi-channel recording devices with wireless power transfer is being implemented with smart devices and arenas for in-sensor analytics. The evolution and detailed comparison of the state of the art in biopotential recording has been described in a later section. 

Biopotential signals, both non-invasive (skin surface) and invasive, have been studied as a means of building bio-electronic medical devices. The central nervous system controls the body and this control can be observed by studying the changes in the peripheral physiological factors such as changes in the heart rate, muscle activity, and breathing. To study these changes, long term monitoring of these physiological signals is necessary\cite{shikano2018simultaneous}. EKG is one of the most widespread diagnostic tools in medicine and the similarity between human and rat EKG\cite{konopelski2016electrocardiography} has permitted the study of various physiological conditions and cardiac diseases \cite{pereira2010noninvasive,zigel2011surface}. 
Along with EKG signals other surface biopotentials such as sEMG and EEG are studied in rats, analysis of these signals is used in sleep studies, epilepsy, locomotive analysis, and effect of spinal cord injuries \cite{keller2018electromyographic,sitnikova2016rhythmic}. 

The study of the brain along with the body is essential in understanding the control mechanisms of the brain on physiology. Sican Liu described a novel neural interface system for simultaneous stimulation and recording of EEG/EMG and ENG (Electroneurogram) signals \cite{liu2019novel}. Along with surface biopotential signals, invasive recording allows for localized, high fidelity signal analysis. Neural biopotential signal analysis is a topic of extensive research in experimental neuroscience, with the aim of improving the quality of life of people with severe sensory and motor disabilities. Wireless neural recording systems have been described in insects, rodents and non-human primates. In rodents particularly, various neural interface systems which include bidirectional communication has been explored\cite{angotzi2014programmable, hampson2009wireless}. Application-specific integrated circuit (ASICs) for neuro-sensing applications has been described for implantable neurosensors\cite{yin2013100,borton2013implantable,biederman2013fully,lee2018implantable}. 

Chronic multi-channel neural recording is a powerful tool in studying dynamic brain function. Multi-electrode arrays permit recording of more than one channel simultaneously enabling neuroscientists to explore different regions of the brain in response to a particular stimulus. Bandwidth constraints limit the number of channels that can be recorded simultaneously resulting in a trade-off between the number of channels that can be simultaneously recorded, power requirements, and the form factor of the device. For example, Borton et al designed an implantable hermetically sealed device that was capable of sending neural signal information via a wireless data link to a receiver placed 1 meter away. This system permitted 7-hours of continuous operation\cite{borton2013implantable}. Chae et al. describes a 128 -channel 6mW wireless neural recording IC with on the fly spike detection for one selected channel. A sequential turn-on method is used to minimize the power requirement \cite{chae2009128}. Similarly, Miranda et al. developed a 32-channel system that can be used for 33-hours continuously but requires two 1200 mAh batteries \cite{miranda2010hermesd}. To achieve a meaningful experimental duration, the power consumption is often >10mW, generally dominated by the communication (radio) power. Thus, it is evident that wireless neural interfaces are power-hungry and there is a need for constant replacement of the batteries or selective channel selection in a chronic setting. To overcome these constraints wirelessly powered neural interfaces were developed, which eliminates the need for constant replacement of the batteries. Implantable devices, in particular, need wirelessly powered devices to reduce the need for a battery at the implant site. Enriched experimental arenas allow for the constant transmission of power facilitating chronic recordings. Yeager et al. developed a wireless neural interface, NeuralWISP capable of sending neural information over a 1-m range \cite{yeager2009neuralwisp}. Lee et. al describes an EnerCage-HC2 to inductively transfer power to a 32-channel implantable neural interface\cite{lee2018implantable}. Wireless power transfer though ensures longer experimental duration, one has to take into account the exposure to high electromagnetic fields along with concerns regarding excessive heat dissipation. Thus, it is evident that neural recordings are limited by size constraints and overall power consumption. This leads to the next advancement in wireless biopotential recording with electro-quasistatic animal body communication. 

\paragraph{}	
\fontsize{11}{12} \textbf{Body Communication Basics}
\paragraph{}

Body communication-based wearable technology has gained prominence over recent times as a communication modality for sending real-time information. 
\begin{figure}[!ht]
\centering
\includegraphics[width=\linewidth]{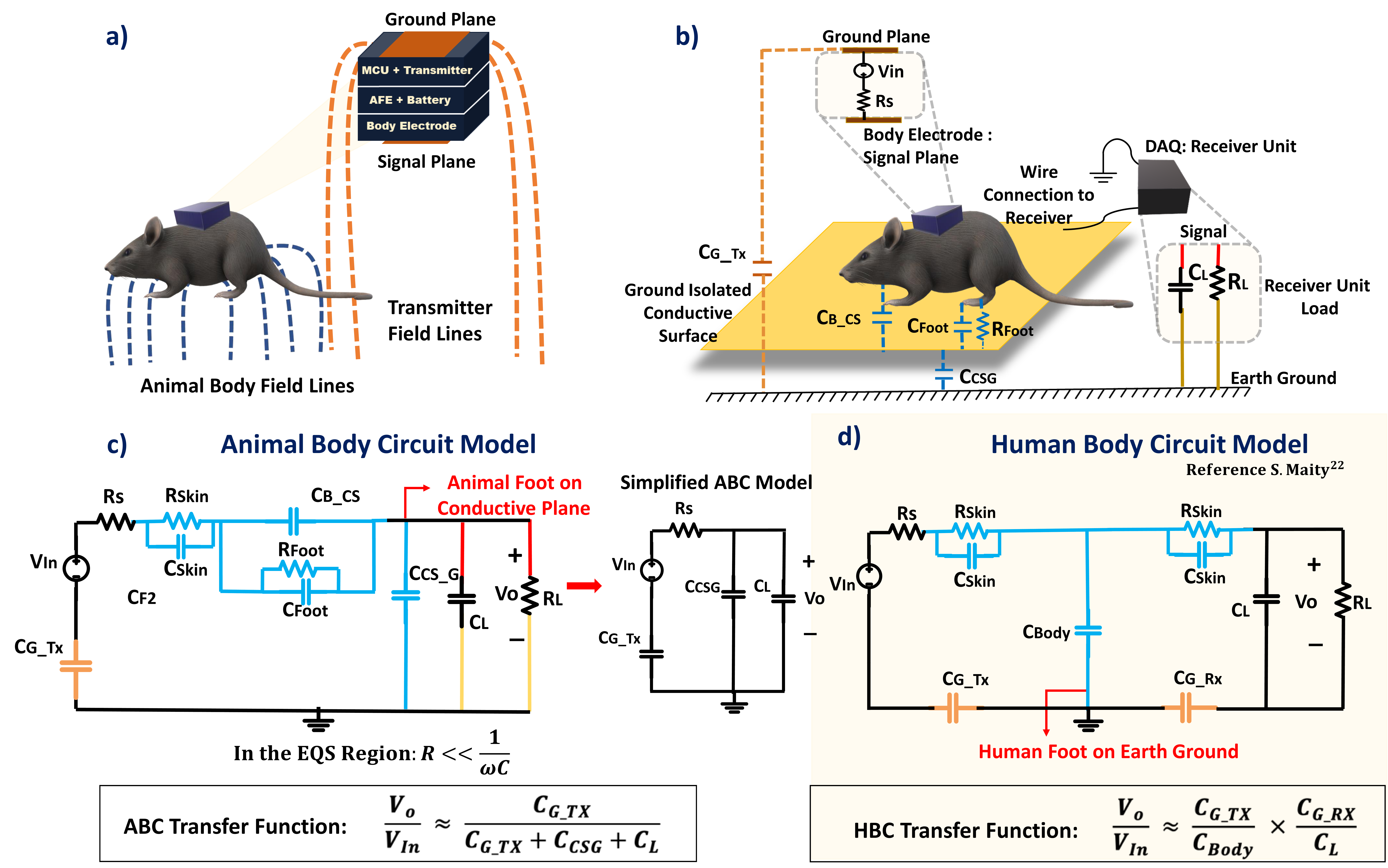}
\caption{Animal Body Communication model: a) Represents the field lines corresponding to the rat body and the transmitter ground plane.  b) The rat body couples to the conductive plane and the associated capacitances are depicted. The conductive plane is ground isolated and forms the capacitance C\(_{CSG}\).  The node consists of the ground plane which couples with the earth's ground to form the capacitive return path. c) The circuit model associated with the experimental setup is shown in the figure. The simplified model of the animal body communication circuit shows that the output voltage is proportional to the conductive plane to ground capacitance C\(_{CSG}\), return path capacitance C\(_{G\_TX}\), and load capacitance C\(_{L}\).  d) Simplified Human Body Communication circuit model and transfer function as depicted by S. Maity \cite{maity2018bio}. }
\label{fig:Figure2_Circuits}
\end{figure}

Recent advances in using the human body as a channel for bio-physical communication has resulted in an energy-efficient information exchange modality. HBC was first proposed as a method to connect devices on a Personal Area Network (PAN) by Zimmerman et al. \cite{ zimmerman1996personal}, using a capacitively coupled HBC model where the return path is formed by the electrode to ground capacitance. The transmitter capacitively couples the signal into the human body which is then picked up at the receiver end. Galvanic coupling-based HBC  introduced by Wegmueller et al.\cite{wegmueller2009signal}, the signal is applied and received differentially by two electrodes respectively. HBC utilizes the conductivity of the human body for a low transmission loss, high-efficiency transmission modality making it ideal for energy-constrained devices. Traditional wireless body area networks (WBAN) use EM signals that radiates outside the body all around us, resulting in only a fraction of the energy being received. This radiative nature and high frequencies in WBANs are typically high energy and of the order of 10nJ/bit\cite{sen2016context}.  
Now, if the body's conductivity is used, it provides a low loss broadband channel that is private (the full bandwidth is available for communication). This low loss and wide bandwidth availability along with the low-frequency operation results in ultra-low power body communication at 415 nW\cite{maity2020415} as well as very low energy communication at 6.3pJ/bit\cite{maity2019bodywire}. Low-frequency HBC was not widely adopted due to the high loss at these frequencies because of resistive (50 $\Omega$) termination\cite{bae2012signal}. Recently we demonstrated, by using capacitive termination, the loss in the EQS region is reduced by a factor of >100, making it usable\cite{das2019enabling,maity2019bodywire}. The first bio-physical model for EQS-HBC was developed by S.Maity\cite{maity2018bio} and a detailed understanding of the forward path\cite{maity2018characterization} and return path\cite{nath2019towards} was described. EQS-HBC is presently the most promising low-power, low-frequency communication alternative for WBAN. It has also been shown that the EQS-HBC adheres to the set safety standards\cite{safety}.

The state of the art in body communication has been restricted to human body communication. In this work we propose to utilize the recent developments in the concept of body communication and apply it to the animal body for biopotential and neural recordings, drastically reducing the size, weight, area, and power of the device. We propose a capacitive termination EQS communication from a sensing node on the rat's body and also device an experimental arena to pick up these EQS signals most efficiently. This form of communication utilizes electro-quasistatic transmission through the conductive layers of the rat below the skin surface. The skin is a high impedance surface while the inner tissue layers are conductive. The transmission of the electro-quasistatic signals through the body with a capacitive return path at frequencies below 1 MHz ensures that the signal is contained within the body. 

\paragraph{}	
\fontsize{11}{12} \textbf{Animal Body Communication - Biophysical Theoretical Model } 
\paragraph{}
As already established, Human Body Communication has been explored as a viable communication model, extending this to an animal body allows for a low loss, efficient channel model, compared to the traditional wireless modalities currently used. Figure \ref{fig:Figure2_Circuits} \textbf{a} and \textbf{b} depicts the concept of Animal Body Communication, the rat body capacitively couples with the signal plane. The transmitter placed on the body of the rat modulates this electric field to transmit OOK (On-Off Keying) sequences corresponding to the sensed biopotential signal. The experimental arena is designed such that the animal moves around on a conductive surface, which is isolated from the earth's ground. This surface picks up the EQS signals coupled onto the animal's body and is received through ground-referenced receiver. Hence, the received voltage is inversely proportional to the capacitance of the signal plane to ground (the less the capacitance the easier it is for the wearable device on the animal to modulate the potential of the animal body and the surface). \begin{figure}[t]
\centering
\includegraphics[width=\textwidth]{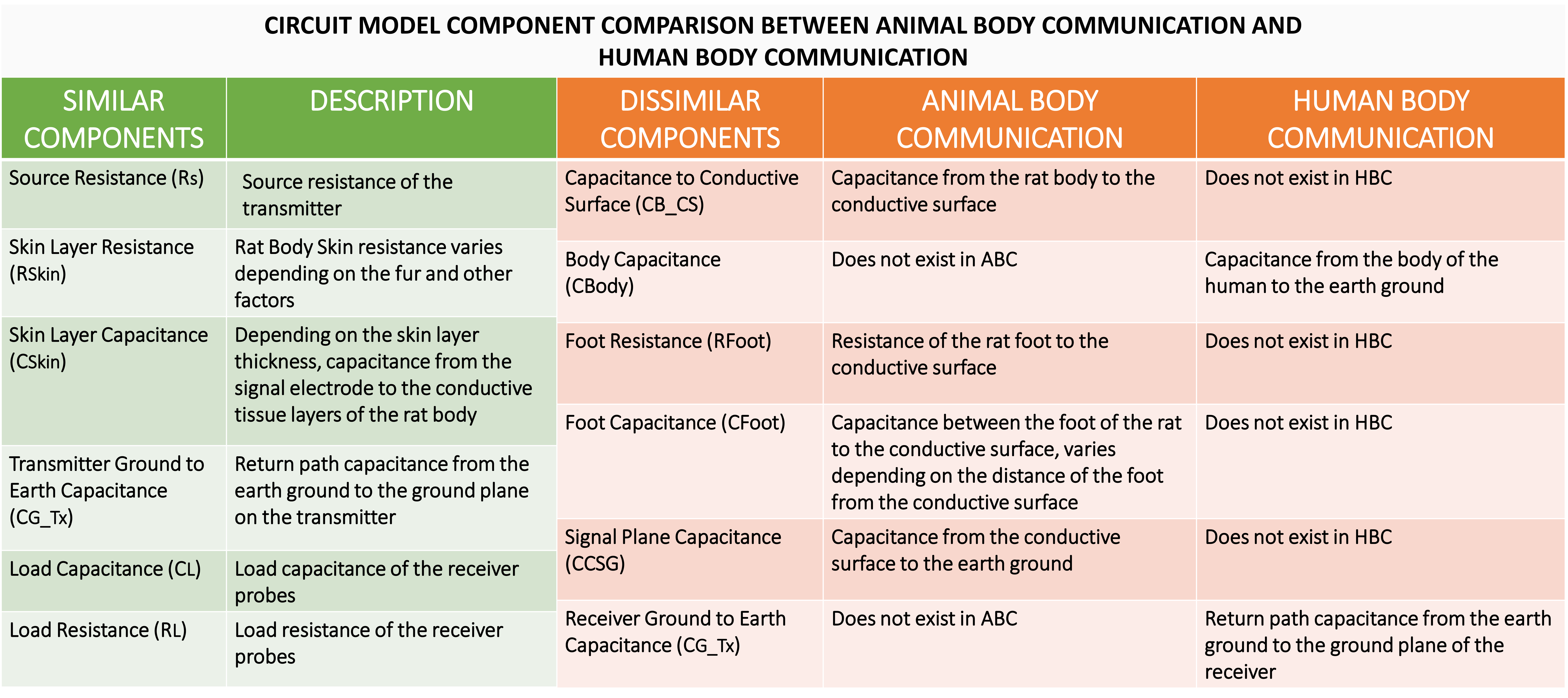}
\caption{Comparison between Animal Body Communication and Human Body Communication circuit components.}
\label{fig:Table}
\end{figure}
The circuit model for Animal Body Communication is described in Figure \ref{fig:Figure2_Circuits} \textbf{c}. At lower frequencies the skin impedance and the series body and foot impedance is negligible compared to the capacitance between the signal plane and ground. Given the operation of ABC in the electro-quasistatic regime, these impedances can be neglected in the computation of the channel loss. From the simplified circuit model, the output voltage \textbf{V\(_{o}\)} and the input voltage \textbf{V\(_{In}\)}  are related as follows:

\begin{ceqn}
\begin{align*}
\textbf{Z\(_{Skin}\)} ,   \textbf{Z\(_{Body}\)} ,  \textbf{Z\(_{Foot }\)} \ll \frac{1}{\textbf{$\omega $ C\(_{CSG }\)} 
}  
\end{align*}
\end{ceqn}

\begin{ceqn}
\begin{align} \label{eq:1}
\frac{\textbf{V\(_{o}\)}}{\textbf{V\(_{In}\)}} = \frac{\textbf{C\(_{G\_TX}\)}}{\textbf{C\(_{G\_TX}\)} +\textbf{C\(_{CSG}\)} + \textbf{C\(_{L}\)}}
\end{align}
\end{ceqn}

In Figure \ref{fig:Figure2_Circuits} \textbf{c} and \textbf{d} we compare the animal body circuit model along with the established human body circuit model. The output voltage is a function of \textbf{C\(_{CSG}\)} as shown in equation \ref{eq:1} for ABC while it a function on  \textbf{C\(_{Body}\)} in case of HBC. Body communication-based systems heavily depend on the body surface and ground sizes. S. Maity describes the Bio-Physical Model for HBC\cite{maity2018bio},  figure \ref{fig:Table} compares the ABC model with the HBC model. Traditional HBC systems have the human body connected to a transmitter and a receiver placed on a different part of the body. The human body has a much larger surface area when compared to an animal. In this ABC setup, the sensor node is placed on the body of the rat, while the receiver is a large conductive plane. This large conductive plane ensures that the movement of the rat is not restricted, and data can be continuously recorded.  In contrast, in human body communication, the body is on the earth's ground and there exists a trunk path to ground.  Due to this, the output voltage is affected by the body capacitance, unlike in the animal body setup. Figure \ref{fig:Table} illustrates the key components of HBC and ABC. The capacitance of the body varies from ABC and HBC due to the fact that the ABC channel model consists of the additional conductive surface on which the rat is free to move. Another important component is the rat foot impedance, in ABC the rat's feet rest on the conductive surface.  \textbf{C\(_{Foot}\)} and \textbf{R\(_{Foot}\)} change depending on the position of the rat's foot on the conductive plane. 

 In the human model, the received signal is collected from the body surface itself, thus the output voltage depends on the capacitive return path of both the transmitter and the receiver. In the ABC model, the conductive surface is ground isolated and connected to an oscilloscope which acts as the receiving unit. The transmitter couples to the floating body and the return path capacitance \textbf{C\(_{G\_TX}\)} from the earth's ground plane to the transmitter ground plane completes the loop, allowing for signal transmission. The receiver in ABC is the oscilloscope signal probe, which can be modeled as the load capacitance \textbf{C\(_{L}\)} in parallel with the load resistance \textbf{R\(_{L}\)}. This oscilloscope is earth ground referenced and hence eliminates the capacitive return path of HBC. The low-loss in ABC coupled with low-carrier frequency communication (as a wire) enables ABC power consumption to be much lower when compared to wireless communication modalities such as Bluetooth. This reduced power enables longer duration experiments with small form factor devices.

\section*{Results}

Animal Body Communication was explored as a new modality for the transmission of biopotential signals. The sensing and transmitting devices are built using off the shelf components and consist of a communication module, a processing module, a power source, and an interface to connect it to the rat body.  Surface electrodes are placed on the skin surface of the rat, after employing appropriate skin preparation techniques and then connecting the electrodes to the front end of the device. Biopotential information is sensed and modulated for transmission, simultaneously transmitting the signal over Bluetooth and through the body of the rat as Animal Body Communication. Bluetooth has long been used as a wireless communication modality and widely cited in literature as a means to transmit biopotential information. In this work, we use this gold standard of communication to compare the biopotential information received from the ABC transmitter and Bluetooth module. A correlation analysis is performed to compare both signals.
Experiments were performed on a rat to prove the feasibility of Animal Body Communication.

 \paragraph{}
\fontsize{11}{12} \textbf{ Animal Body Communication Experimental Setup}
\paragraph{}

The Animal Body Communication setup is tested on Sprague Dawley rats, experiments were performed on anesthetized rats. In this study, capacitive coupling is used as a means to achieve Animal Body Communication. The details of the sensor node are described in the methods section. 

\begin{figure}[!ht]
\centering
\includegraphics[width= \textwidth]{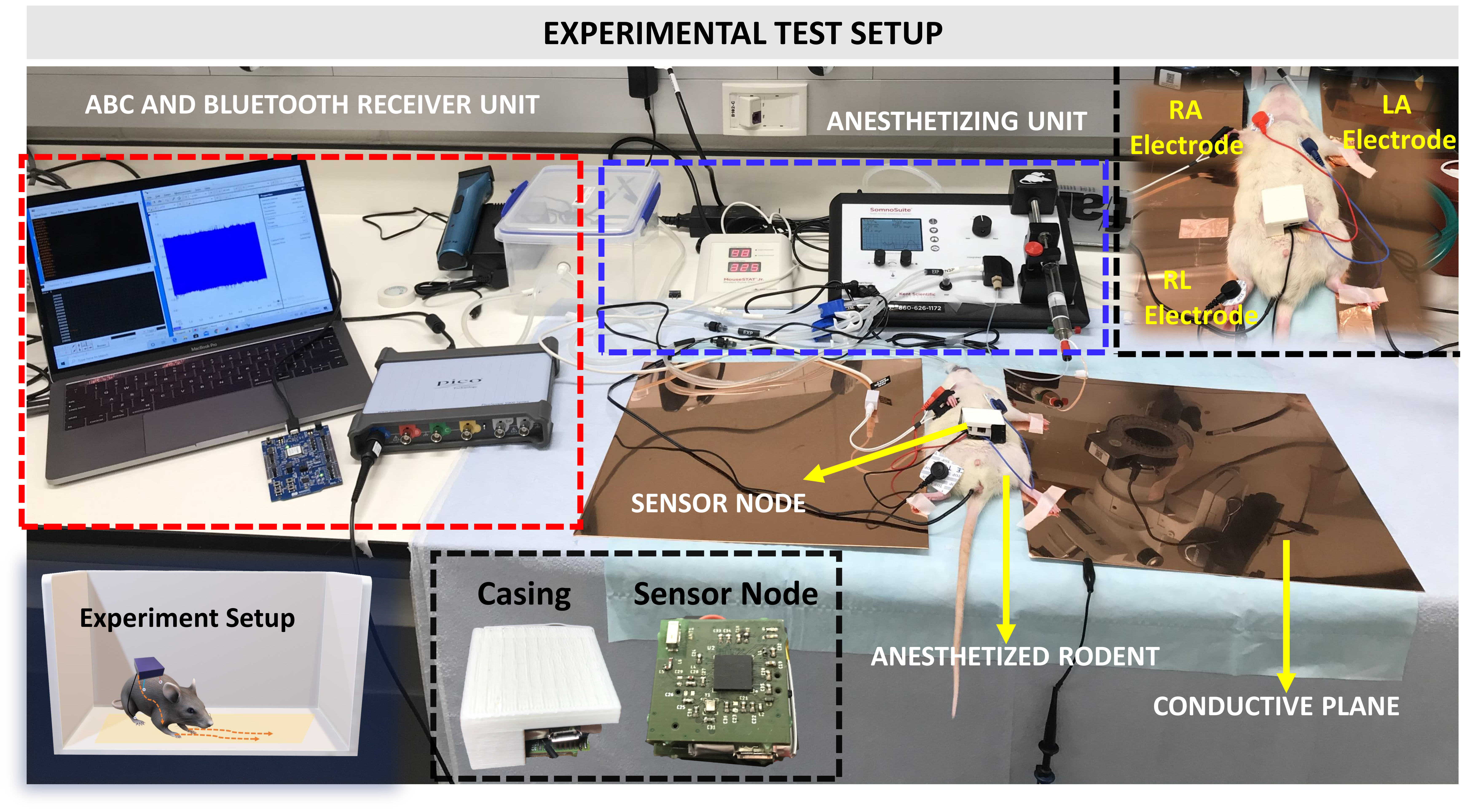}

\caption{Experimental Test Setup: The sensor node is placed on the rat skin surface and connected to the surface electrodes (Right Arm (RA), Left Arm (LA), and Right Leg (RL) for EKG sensing. The rat's feet are taped on a conductive copper plate, the plate is then connected to a receiver for ABC transmission. The Bluetooth receiver and the ABC receiver are connected to the computer for signal acquisition and processing. The setup also consists of the anesthetizing unit which delivers the anesthetizing drug and oxygen to rat.}
\label{fig: CompleteSetup}
\end{figure}

	Anesthetized rats are placed on a non-conductive surface, the sensor node, in a casing, is placed on the rat skin surface and patch connectors are used to connect to the surface electrodes. The feet of the rat are placed on a conductive copper plate, signals are acquired using the sensing unit, then transmitted via Bluetooth to a receiver connected to a computer as shown in Figure \ref{fig: CompleteSetup}. Only the feet are connected to the conductive plane while leaving the body on a non-conductive surface. This depicts a case when the rat moves in a cage with only the feet on the bottom plane. ABC happens through the transmission of OOK sequences from the node through the body, to the conductive copper plate. These signals are picked up using an oscilloscope connected to the conductive plane. The oscilloscope signal probe is connected to the conductive plane while the ground probe is left floating. EKG signals are acquired using a three-electrode setup with the electrodes placed on the Right Arm (RA), Left Arm (LA), and Right Leg (RL). The RL serves as the right leg drive, common to EKG recording systems. Additional monitoring systems such as the anesthetizing setup and body vital measurement systems are present in this experimental setup not part of the communication setup. This setup aims to mimic the setup as described in Figure \ref{fig:Figure1ABC}. The copper plates act as the conductive surface which in an awake recording setup will form the base on which the rat is free to move.
    
\paragraph{}
 \fontsize{11}{12} \textbf{ Time Division Multiplexing}
\paragraph{}

Biopotential signal measurements require the body to be grounded to improve the CMR of the entire system. Grounding the body eliminated the floating nature which is essential for body communication. Thus, to sense and transmit biopotential signals, time-division multiplexing is used. Such multiplexing between each sensing cycle and transmission cycle ensures that surface biopotentials can be sensed accurately and also transmitted via body communication. 

 \begin{figure}[!ht]
\centering
\includegraphics[width= \textwidth]{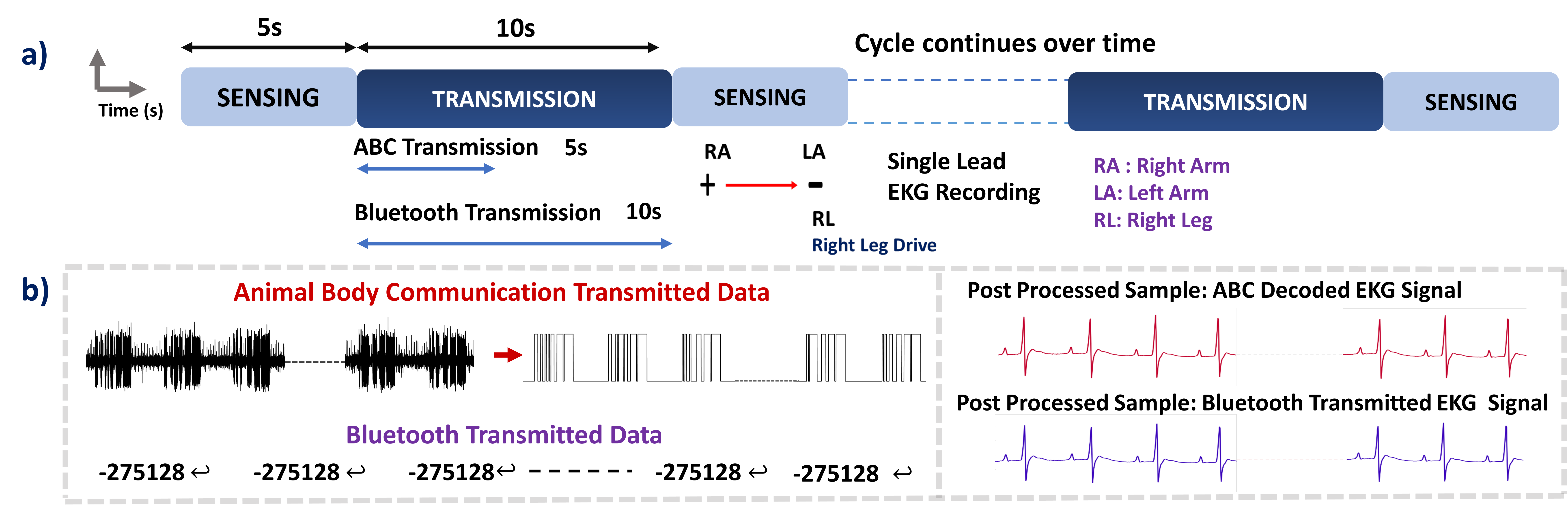}
\caption{Time multiplexed sensing and transmission cycles: a) Sensing cycles consists of acquiring the single-lead ECG signal using three electrodes, Right Arm, Left Arm and Right Leg. ABC Transmission starts at the end of the sensing cycle and has the same duration as the sensing cycle. Bluetooth transmission duration is almost twice the ABC transmission duration. Upon the completion of both transmissions, the sensing cycle restarts, and this cycle repeats.  b) Post-processing steps on ABC transmitted sequence and Bluetooth transmitted sequence transmission}
\label{fig:Transmission}
\end{figure}

In the event of simultaneous sensing and transmission, given that the transmitter is placed on the surface of the body, the sensing electrodes pick up the OOK sequences used in the transmission, resulting in a corrupted sensed signal. To avoid this, sensing and transmission are time multiplexed. This technique is critical for body communication with surface biopotentials. Figure \ref{fig:Transmission}\textbf{a }describes the time multiplexing cycles, data is sensed for a period of 5s followed by the transmission for 10s. The transmission of ABC and Bluetooth occurs simultaneously, however Bluetooth sequences take longer to transmit due to packet constraints resulting in a longer transmission time as compared to the sensing time. Following the transmission cycle, the sensing cycle repeats. ABC data is sent as OOK sequences which are then demodulated and decoded to retrieve the EKG sample as shown in Figure \ref{fig:Transmission}\textbf{b}. Bluetooth samples are transmitted as characters corresponding to the ADC codes, which are then converted to corresponding samples to compare with the transmitted ABC signal. 

\paragraph{}
\fontsize{11}{12} \textbf{ Time Domain Correlational Analysis on Acquired EKG Signal}
\paragraph{}
 EKG signals are chosen for testing the animal body communication setup. The experiment was conducted on a total of 8 Rats over 2 months. This current set-up ensures continuous synchronized transmission of the biopotential signal from both the Bluetooth module and the ABC transmitter. 

 \begin{figure}[!ht]
\centering
\includegraphics[width= \textwidth]{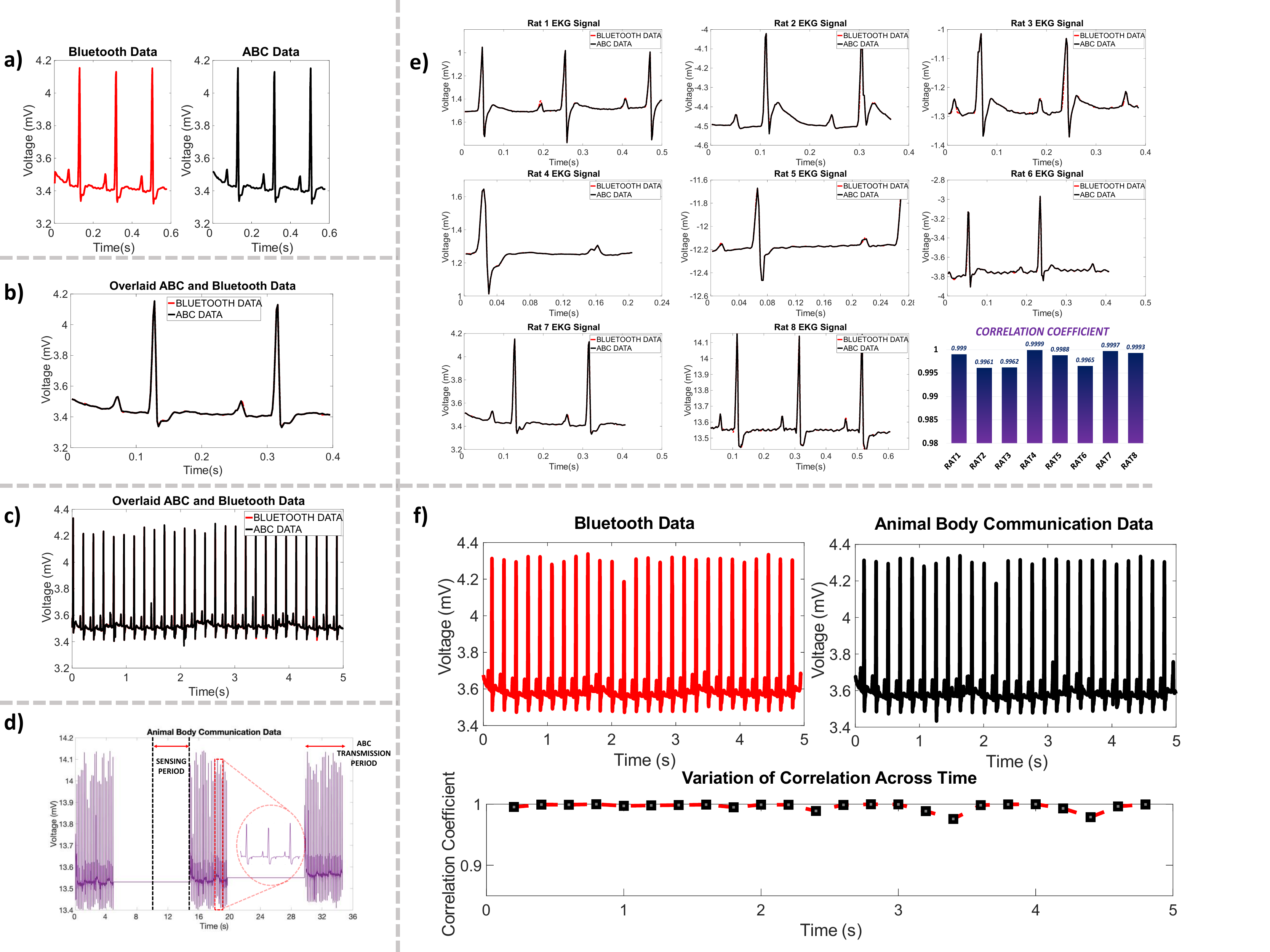}
\caption{ Rat Electrocardiogram (EKG) Analysis: a) Bluetooth and ABC transmitted signal.  b) Overlaid Bluetooth and ABC transmitted signal data, depicting overlap of the EKG peaks. c) Complete 5s Bluetooth and ABC transmitted EKG Signal. d) Time multiplexed ABC transmission cycles, 5s transmission time followed by a 10s wait time to allow for Bluetooth transmission and next cycle sensing. e) Overlaid plots of EKG Signals from eight rats with correlational analysis between Bluetooth and ABC transmitted signals. f) Time varying correlational analysis of one 5s sensing cycle of rat EKG signal.}
\label{fig:Results_New}
\end{figure}

As mentioned before, the signals are time-multiplexed allowing Animal Body Communication. The EKG signal is sensed for a period of 5s followed by simultaneous transmission of ABC and Bluetooth. Figure \ref{fig:Results_New} shows the EKG sample comparison, \ref{fig:Results_New}\textbf{a} shows the Bluetooth and the ABC EKG data for a period of 0.6s, these two signals are overlaid in \ref{fig:Results_New}\textbf{b}, the PQRST peaks of the characteristic EKG signal align, similarly, the data is compared for all 8 rats and correlation accuracies across each trial is depicted in \ref{fig:Results_New}\textbf{e}. The correlation accuracy for all the rats was seen to be > 99.5 \%. In \ref{fig:Results_New}\textbf{c}, we can see the complete overlaid 5s sample. Time multiplexing results in an ABC transmission period followed by a wait time for the completion of the Bluetooth transmission and sensing. \ref{fig:Results_New}\textbf{d} depicts this time-multiplexed ABC data, with this cycle being continuous.  \ref{fig:Results_New}\textbf{f} depicts the variation of correlation across the entire 5s window, the correlation between Bluetooth and ABC is approximately 1 throughout the 5s window depicting a reliable transmission system.

\paragraph{}
\fontsize{11}{12} \textbf{ Effect of Distance of Foot from Conductive Surface to Received ABC Signal}
\paragraph{}

\begin{figure}[t]
\centering
\includegraphics[width=\textwidth ]{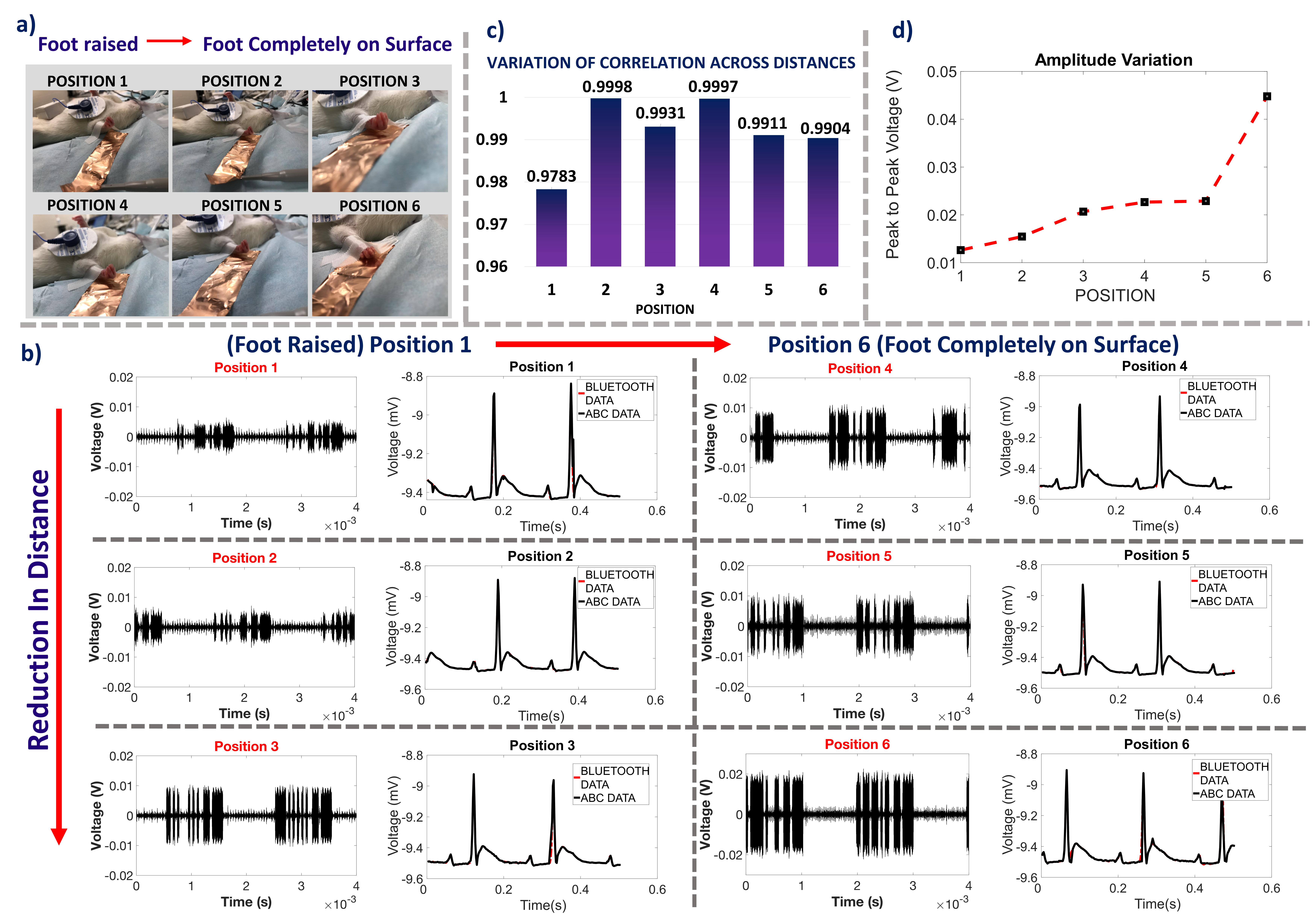}
\caption{ Effect of distance on received signal strength: a) Experimental setup to measure the effect of distance variation on received OOK signal; In position 6, the rat foot is closest to the conductive surface and in position 1 the foot is furthest from the conductive surface.  b) Received OOK signals of different distance variations and the decoded signals for the different distances c) Correlation between Bluetooth and ABC received signals as a function of distance, each depicting correlation accuracy >97\%. d) Variation in the amplitude of the received signal as a function of distance.}
\label{fig:DistanceVariations}
\end{figure}

A key component of animal body communication is the dependence of the received signal on body resistance and capacitance. Variation of the distance of the foot from the conductive surface changes the magnitude of the received signal which then tests the robustness of the system. Experimental analysis with only one foot on a conductive surface with varying distances shows that even with the foot raised, OOK sequences can be picked up from the conductive strip. It is evident that as the distance from the conductive surface reduces, the amplitude of the coupled signal increases. However, even at large distances, though the signal amplitude is lower, the received Bluetooth and ABC signal can be decoded and display >97\% correlation. When the rat foot is raised above the conductive surface, the resistance becomes infinite, but the capacitance between the rat foot and body, to the conductive surface exists as shown in Figure \ref{fig:Figure2_Circuits} and this ensures the necessary path for transmission of the signal. Since body communication works on capacitive coupling, even without complete contact with the conductive surface, the OOK sequences couple to the conductive surface. It is highly unlikely that the rat would have all feet raised above the conductive plane, for a long time. In the event of improper contact with the conductive surface or when the rat jumps, it is shown that the signals can still be received on the conductive plane and can be successfully decoded. For short burst errors, the implementation of error-correcting codes can ensure robust transmission. In Figure \ref{fig:DistanceVariations} the variation of the distance of the rat foot from the conductive surface is shown, position 1 is furthest away from the conductive strip, while in position 6, the rat foot is completed taped on the conductive surface. The amplitude of the received signal increases with the reduction in distance however in all cases it can be seen that the sequences can be decoded and all show high correlation accuracies with Bluetooth.

\section*{Discussions}

Capacitive coupling from the transmitter ground plane to the earth's ground ensures the return path necessary for animal body communication. 
\begin{figure}[!ht]
\centering
\includegraphics[width=\linewidth]{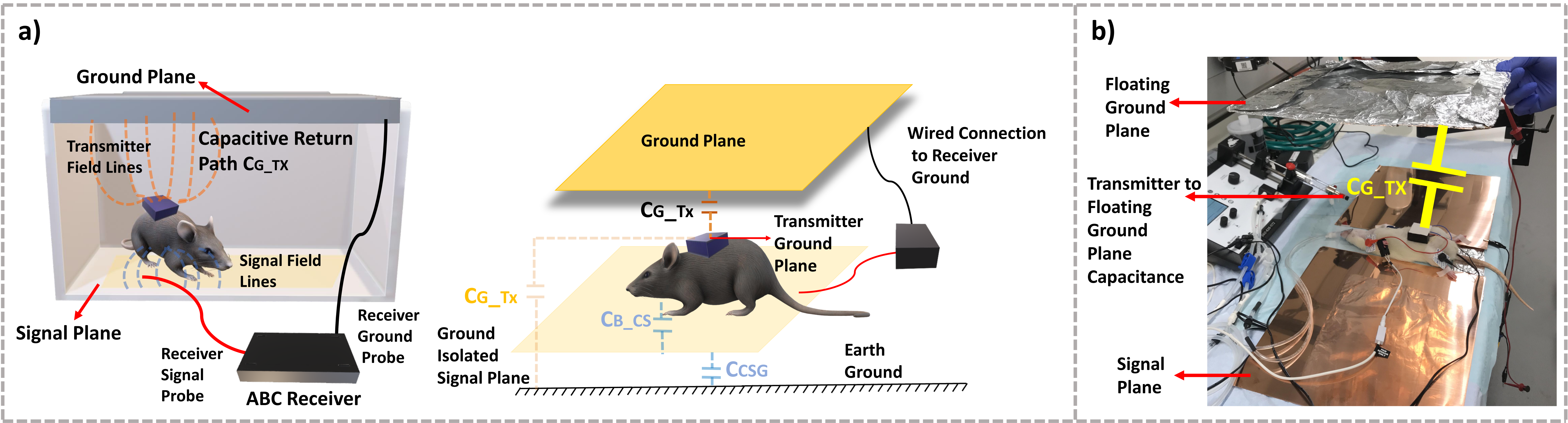}
\caption{Effect of an External Ground Plane: a) Experimental setup depicting the electric field lines from the transmitter ground plane to the external ground plane; Capacitive return path from the transmitter ground to the external ground plane. b) \textit{In vivo} test setup with a hand-held external ground plane. The external ground plane is placed above the rat body similar to the experimental setup, mirroring the top of a rat cage. }
\label{fig:Discussion}
\end{figure}
The presence of a large conductive signal plate prevents the existence of such a capacitive path to ground. The addition of a conductive plane connected to the receiver ground placed above the rat body provides the necessary return path. The transmitter ground plane along with this floating ground plane forms the capacitance  \textbf{C\(_{G\_TX}\)}. In the setup with a rat cage, as shown in Figure \ref{fig:Discussion}\textbf{b} the top and bottom surfaces of the rat cage are made conductive, with the top plate connected to the receiver ground, while the bottom plate, which acts as the signal plane is connected to the signal probe of the receiver. During \textit{in vivo} tests, the ground plane consisted of a hand-held conductive plane above the anesthetized rat body. Only the feet of the rat are connected to the signal plane, with a slot in the conductive plane to allow for the placement of the rat. Figure \ref{fig:Discussion}\textbf{a} describes the need for the addition of the conductive ground plane in a model rat cage. Similar to Figure \ref{fig:Figure2_Circuits}\textbf{b}, the capacitive coupling from the device ground plane to the external ground plane provides the necessary return path. The sensor node placed on the body of the rat has the transmitter ground plane on the top surface and the signal electrode touches the body of the rat. The addition of this floating ground plane allows for the use of a large signal plane, providing a larger experimental arena for the rat to move on without being limited by the loss of the signal return path.

\section*{Conclusion}
\paragraph{}

To conclude, in this work we demonstrate a novel communication modality in the animal studies domain and show the use of the animal body as a communication channel.
\begin{figure}[!h]
\centering
\includegraphics[width=\textwidth ]{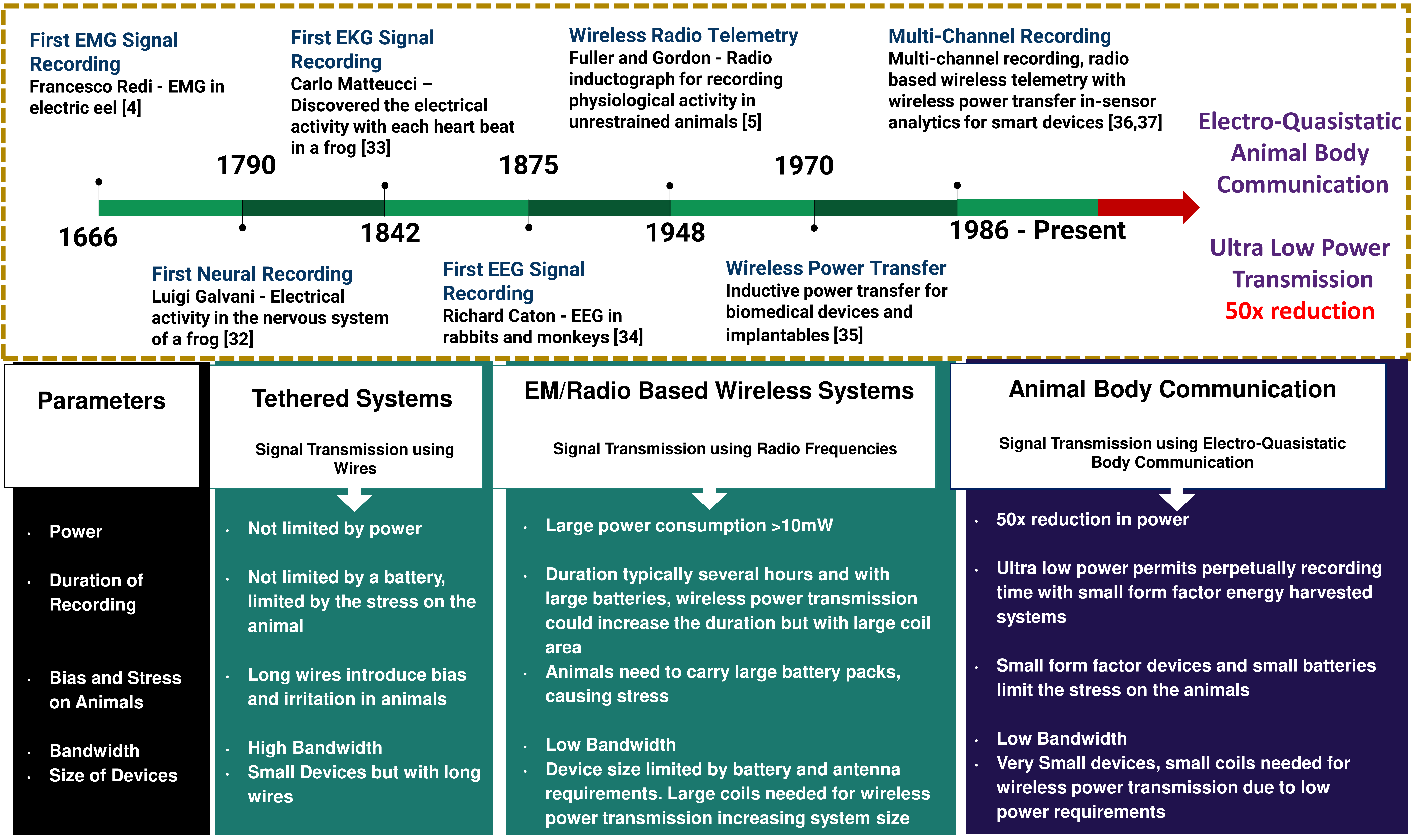}
\caption{Evolution of animal biopotential recordings \cite{reaz2006techniques,piccolino1997luigi,alghatrif2012brief,caton1875electric,fuller1948radio,shadid2018literature,sodagar2009wireless,najafi1986implantable}; Comparison of tethered, traditional wireless systems and Animal Body Communication.}
\label{fig:Evolution}
\end{figure}
Biopotential signals were acquired from the rat and transmitted using Animal Body Communication. The theory and channel model for animal body communication was developed and a custom-designed sensor node was built and tested \textit{in vivo}. The correlation accuracy between standard wireless transmission systems and ABC was found to be > 99 \% in these tests. The power consumption for Bluetooth transmission was observed to be 29.5 mW, while the power consumption for ABC transmission was found to be 0.5 mW. This depicts a > 50x reduction in power. If a custom-designed IC is built with only ABC transmission, the device size and power can be further significantly reduced, along with the possibility to make these high bandwidth systems. The effect of variation of distance of the foot of the rat from the receiver signal plane was observed and it is clear that reliable signals can be received even with improper contact or raised feet, adding to the reliability of this communication channel. A modified test setup was explored as an additional technique to ensure robust communication. While in this study, EKG was the chosen biopotential, it can be extended to neural signal acquisition and transmission, where low power communication modalities are essential. In Figure \ref{fig:Evolution} the evolution of animal biopotential recording was studied, the key differences between tethered, wireless and EQS-ABC was compared and it was found that EQS-ABC can prove to be the next advancement in this domain, allowing for an ultra-low power, efficient channel model.

\section*{Methods}
\paragraph{}	

\begin{figure}[!h]
\centering
\includegraphics[width=\linewidth]{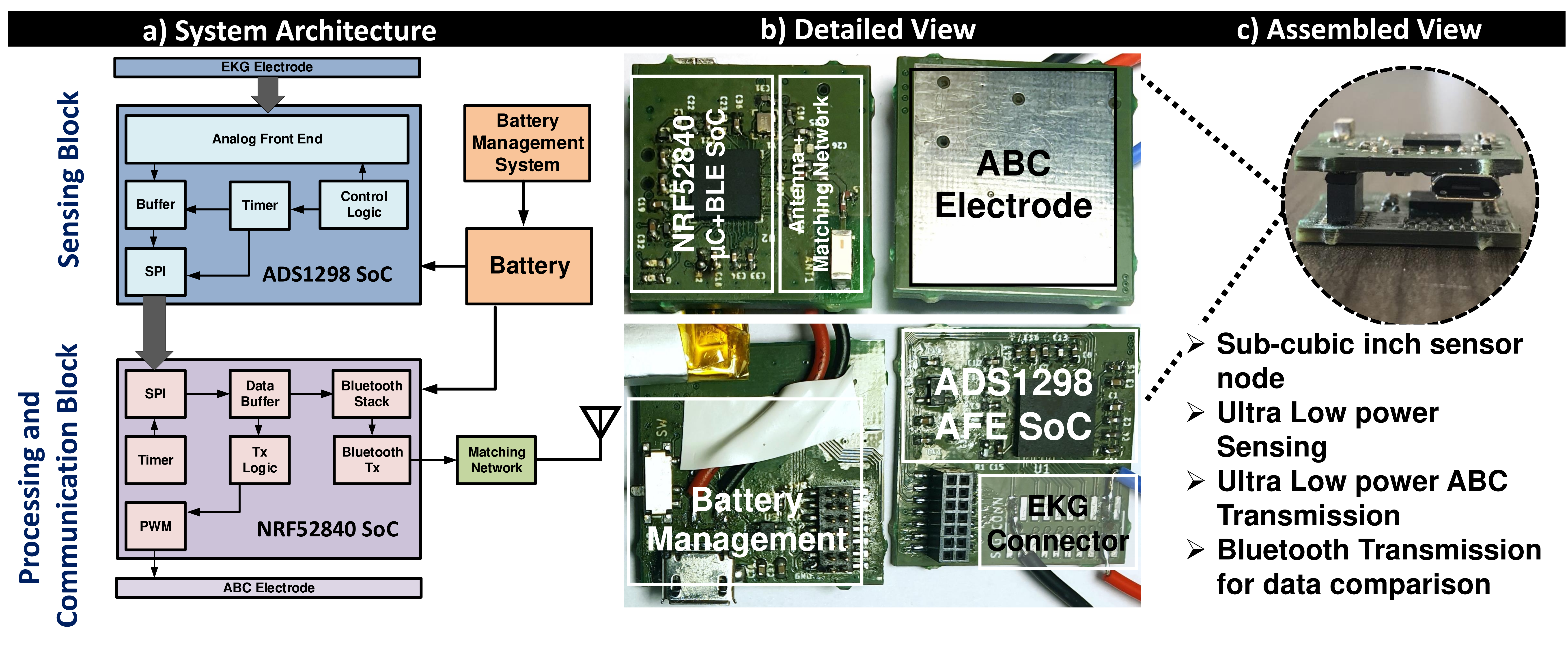}
\caption{System Architecture of the custom-built node for biopotential acquisition through animal body communication and Bluetooth Low Energy; a) Block diagram of the custom-built node, b) Functional blocks depicted on the actual device, c) Custom node after stacking.}
\label{fig:System_Architecture}
\end{figure}

\fontsize{11}{12} \textbf{ System Architecture } 
\paragraph{}
Size, weight, area, and power consumption of wireless recording devices have the potential to significantly affect animal behavior and compromise the quality and length of recordings, thereby hindering scientific studies. Overcoming these obstacles formed the core design objectives for the custom node for the acquisition of biopotential signals and wireless transmission of data and resulted in the following initial specifications. Physical dimensions were constrained to one cubic inch, which is sufficiently small to be placed on a rodent and large enough to house the various components. The net weight and power consumption were capped at 50g, and 50 mW respectively. This posed a significant challenge since the analog front end for sensing, micro-controller for computing, wireless communication for comparison purposes, power management, and animal body communication had to be miniaturized and integrated into the device while meeting the power budget.

The system architecture as shown in Figure \ref{fig:System_Architecture}\textbf{a} can be broadly divided into three blocks, the custom-wireless signal acquisition node, the Bluetooth receiver connected to the data logging system (computer), and the animal body communication receiver. The custom node consisted of two vertically stacked custom-designed printed circuit boards (PCB) which were populated with commercially available integrated circuits and discrete components. The top board in the stack contained the micro-controller and Bluetooth System on Chip (SoC), along with the antenna and matching network on the top layer. The bottom layer consisted of the power management system and charging connector. The analog front end was housed on the top layer of the bottom stack, with the bottom layer serving as the electrode for animal body communication. The detailed view and the assembled view of the sensor node is shown in as shown in Figure \ref{fig:System_Architecture}\textbf{b} and \ref{fig:System_Architecture}\textbf{c} respectively.

A System on Chip (NRF52840, Nordic Semiconductors) which integrates an ARM Cortex-M4F micro-controller and a Bluetooth 5.0 transceiver was selected to form the core of the node since it would minimize the device footprint and power consumption. The on board 1MB flash memory and 256KB RAM was sufficiently large to store the sampled signals and implement in-sensor analytics in the future. Power efficiency was further improved by utilizing the on-chip DC-DC converters.

The custom node collected the EKG signals from a zero-insertion force connector placed on the PCB. Signal conditioning and sampling of the EKG signal was performed by another SoC (ADS1298, Texas Instruments). This analog front-end chip incorporates a programmable gain differential amplifier and right-leg drive generation for conditioning EKG signals, which were subsequently sampled at 500Hz by a 24-bit analog to digital converter. The SoC was programmed to optimize signal acquisition quality and power consumption. The sampled signals were sent to the micro-controller through an on-chip Serial Peripheral Interface.

The sampled data was stored in a buffer in the micro-controller until the transmission window started. The samples were then converted to characters and transmitted as a string over Bluetooth after adding delimiters to differentiate between subsequent samples. For Animal Body Communication, the sample was transmitted in its original 24-bit binary integer form after creating packets by adding two bits (binary 1) at the start and end of the sample. Each bit in ABC was represented by on-off keying, wherein a 500kHz, 50\% duty cycle square wave was turned on (binary 1) or off (binary 0). ABC data was transmitted at 25Kbps, which was significantly lower than the minimum required Bluetooth bandwidth of 45Kbps, which excludes the overhead added by the Bluetooth stack.

The custom-designed node was packaged in a 3D-printed housing of dimensions 25mm x 25mm x 10mm, which is equivalent to 0.39 cubic inches. It had a net weight of 20g and average power consumption of 28.5 mW (with Bluetooth transmission for data comparison purposes) which resulted in approximately 20 hours of battery life. This is 19 times smaller and has more than twice the battery life when compared to a commercial wireless unit (Bio-Radio). We expect a much longer lifetime when the Bluetooth transmission is turned off and only ABC transmission is turned on. The power required for sensing is typically orders of magnitude lower than the power required for communication, thus the system power is dominated by this communication power. The ABC transmission power is 50x lower when compared to the Bluetooth transmission power and this translates into an order of magnitude improvement in the device lifetime and reduction in the battery size.

The Bluetooth receiver was essentially another NRF52840 SoC connected via USB to the data logging system, which in this case was a computer. This setup was used instead of the inbuilt Bluetooth device of the computer since it would be easier to collate data from multiple transmitters. 

The conductive signal plane is connected to the high impedance receiver probe. A computer-based oscilloscope, by Pico Technologies, was used as the ABC receiver. The OOK sequences are sampled at 3.9 MSamples/s and collected for post-processing. 

\paragraph{}
\fontsize{11}{12} \textbf{ Signal Processing }
\paragraph{}

OOK sequences collected from the ABC receiver are sent to a computer for processing. Signals are first band-passed between 400kHz to 600 kHz with  80 dB attenuation software filters.  Filtered sequences are demodulated using envelop detection and thresholding. Sequences are then decoded using the start and stop bit followed by software error correction. Bluetooth sequences in the form of ADC codes are converted to corresponding voltage values and compared to the received ABC signals. 
\paragraph{}
\fontsize{11}{12} \textbf{ Communication Protocols}
\begin{itemize}
\item Time Multiplexed Data \\
As discussed earlier, a requisite for animal body communication especially while recording surface biopotential signals is the need to time multiplex the sensing and transmission periods.

\item Error-Correcting Algorithms\\
There is a possibility to bring in redundancy into the communication channel to ensure the robustness of this communication modality. We have shown that if the rat foot is lifted from the conductive surface, the received signal can still be picked up by the receiver. The goal of this paper is to ensure that long term recordings of freely moving animals can be obtained. To ensure that there is a successful transmission of data, error-correcting algorithms become a necessity. \\
Bi-modular Redundancy can be introduced by repeating packets over time. In the event of a jump or signal drop, repeated packets ensure that the signal information is faithfully transmitted. This technique reduces the data rate due to the added redundancy.\\Block Codes a common error-correcting technique of encoding the data in blocks, such that the code is a linear combination of the message and parity bits in a linear block code. 
\end{itemize}

\paragraph{}
\fontsize{11}{12} \textbf{ Surgery }
\paragraph{}

All surgical procedures were performed under aseptic conditions at Purdue Animal Facility. 5\% Isoflurane gas and oxygen were used to anesthetize the rat in an induction chamber, followed by a continuous flow of 2.5 \% Isoflurane gas with oxygen delivered through a nose cone. The dosage of Isoflurane and the flow of oxygen is continuously monitored to ensure that the rat does not respond to the toe pinch while still maintaining a steady breathing rhythm and observable pink extremities. A heating pad is placed below the rat to maintain the body temperature and lubricating drops are added to the eyes of the rat to prevent drying. The skin surface is shaved and cleaned for the placement of the surface electrodes. The device is placed on a shaved surface on the belly of the rat with the signal plane touching the skin surface. The surface electrodes are connected to the device using patch connectors.

All procedures were approved by the Institutional Animal Care and Use Committee (IACUC) and all experiments were performed in accordance with the Guide for the Care and Use of Laboratory Animals. The experiments were closely monitored and reviewed by Purdue Animal Care and Use Committee (PACUC).

\paragraph{}

\bibliography{REF}

\section*{Acknowledgements}
\paragraph{}

This work was supported by National Science Foundation CAREER Award (ECCS 1944602), Air Force Office of Scientific Research YIP Award under Grant FA9550-17-1-0450 and by the NIH Stimulating Peripheral Activity to Relieve Conditions (SPARC) program (OT2OD023847 and OT2OD028183). (The contents of this article do not necessarily represent the official views of the NIH). The authors would like to thank Dr. Shovan Maity, graduated PhD student, Shramana Chakraborty, M S Student, Jongcheon Lim, Umm E Hani Abdullah, David Yang, Arunashish Datta, Debayan Das, Mayukh Nath, PhD Students as well as Visiting Scholar Gargi Bhattacharya at Purdue University for their immense co-operation and support during the experiments.

\section*{Author Contributions}
\paragraph{}

S. Sriram, S. Avlani, M P. Ward, S. Sen,  conceived the idea, S. Avlani and S. Sriram designed the ABC sensor node. S. Sriram conducted the theoretical analysis and performed the experiments under the guidance of M P. Ward and S. Sen. All authors contributed to the drafting of this manuscript and have read and approved the final version of the manuscript.

\section*{Additional Information}
\paragraph{}

\textbf{Competing Interests:} The authors declare no competing interests.

\end{document}